\documentclass[secnumarabic,twocolumn, amssymb, nobibnotes, aps, pra, superscriptaddress]{revtex4-2}
\usepackage{graphicx}
\usepackage{float}
\usepackage{amsmath}
\usepackage{physics}
\usepackage{hyperref}
\usepackage[all]{hypcap}
\hypersetup{
    colorlinks=true,
    linkcolor=blue,
    filecolor=magenta,
    urlcolor=cyan,
}
\newcommand{\Eq}[1]{Eq.\,(\ref{#1})}
\newcommand{\Fig}[1]{Fig.\,\ref{#1}}

\newcommand{\kmax}{\ensuremath{k_\mathrm{max}}}
\newcommand{\Onlinecite}[1]{Ref.~\cite{#1}}
\newcommand{\addmodes}{\textit{gap modes}}

\usepackage{xr} % cross-references to external documents
\begin{document}
\title{Extending completeness of the eigenmodes of an open system beyond its boundary, for Green's function and scattering-matrix calculations
}
\author{Z. Sztranyovszky}
\affiliation{School of Physics and Astronomy, Cardiff University, Cardiff CF24 3AA, United Kingdom}
\affiliation{School of Chemical Engineering, University of Birmingham, Birmingham B15 2TT, United Kingdom}
\author{W. Langbein}
\author{E. A. Muljarov}
\affiliation{School of Physics and Astronomy, Cardiff University, Cardiff CF24 3AA, United Kingdom}
\date{\today}

\begin{abstract}
    The asymptotic completeness of a set of the eigenmodes of an open system with increasing number of modes enables an accurate calculation of the system response in terms of these modes. Using the exact eigenmodes, such completeness is limited to the interior of the system. Here we show that when the eigenmodes of a target system are obtained by the resonant-state expansion, using the modes of a basis system embedding the target system, the completeness extends beyond the boundary of the target system. We illustrate this by using the Mittag-Leffler series of the Green's function expressed in terms of the eigenmodes, which converges to the correct solution anywhere within the basis system, including the space outside the target system. Importantly, this property allows one to treat pertubations outside the target system and to calculate the scattering cross-section using the boundary conditions for the basis system. Choosing a basis system of spherical geometry, these boundary conditions have simple analytical expressions, allowing for an efficient calculation of the response of the target system, as we demonstrate for a resonator in a form of a finite dielectric cylinder.
\end{abstract}

\maketitle

\paragraph*{Introduction.}
The optical response of an object to an excitation is determined by its morphology, material, and the surrounding medium. For spherical resonators the response can be calculated analytically and it is widely known as Mie scattering \cite{MieAP1908}, while for an object much smaller than the wavelength one can use the quasi-static approximation \cite{BohrenBook2012}, or for a weak scatterer Born's approximation \cite{HulstBook2003}. In more general cases, however, numerical methods are needed to solve Maxwell's equation for the scattered field, such as the finite-difference time-domain approach, or the finite-element method in frequency domain \cite{KahnertJQSRT2003}. While directly solving Maxwell's equation with well established methods can be straightforward, it can also be computationally expensive, particularly when a large number of different excitations need to be considered, such as illumination over a wide range of incidence angles and wavelengths, which can be the case when modelling bright or dark field microscopy  \cite{WangNSA2020,WangNS2022}.

An alternative approach, which emerged in the last decade, expresses the response of the open system to excitation in terms of its resonances, or complex eigenmodes \cite{DoostPRA2012,WeissPRB2018,BaiOE2013,SauvanPRL2013,LobanovPRA2018,MuljarovPRB2016Purcell,LalanneLPR2018,BothSST2021}. These eigenmodes, also referred to as resonant states (RSs) \cite{DoostPRA2012,WeissPRB2018}, or quasi-normal modes \cite{BaiOE2013, SauvanPRL2013}, are a powerful concept as they describe observables, such as transmittance \cite{DoostPRA2012,WeissPRB2018}, optical cross-section \cite{BaiOE2013,LobanovPRA2018}, or Purcell factor \cite{SauvanPRL2013,MuljarovPRB2016Purcell} in a mathematically rigorous and physically intuitive way. Using completeness properties of sets of eigenmodes \cite{LeungPRA1994,LeungJOSAB1996,LeungJPA1997,LeeJOSAB1999,SauvanOE2022}, the observables can be expressed as a Mittag-Leffler (ML) series of the Green's function (GF) \cite{DoostPRA2012,LobanovPRA2018,MuljarovPRB2016Purcell} or of the scattering matrix (S matrix) \cite{WeissPRB2018,BothSST2021}, or they can also appear as a weighted sum over the eigenmodes \cite{BaiOE2013,SauvanPRL2013,LalanneLPR2018}.

The question of completeness of eigenmodes, that is, weather they form a suitable basis for expansion of other functions, has been recently reviewed in detail for open systems in \Onlinecite{SauvanOE2022}. Assuming a homogeneous background, we refer to the inhomogeneous region of space as the resonator. In the frequency domain, the region of completeness is assumed to be the minimal convex volume containing the resonator. This assumption has been verified for analytically solvable systems, such as a slab or a sphere \cite{LeungPRA1994,LeungJOSAB1996,LeungJPA1997,DoostPRA2012,MuljarovPRA2020}. When the eigenmodes of the system are calculated numerically \cite{LalanneLPR2018}, additional numerical modes can appear in the spectrum, for example due to the use of perfectly matched layers in bounded regularized space, or due to the discretization of the differential operator, and including all such modes in the basis of expansion can result in completeness beyond the resonator \cite{YanPRB2018}.

While there have been recent attempts to achieve completeness of the RSs in the region outside the resonator, by applying various methods of their regularization \cite{ColomPRB2018,AbdelrahmanOSAC2018,KristensenAOP2020,DezfouliPRB2018,FrankePRA2023}, the completeness of the true, physical RSs is limited to the interior of the resonator. In this Letter, we show that
using the resonant-state expansion (RSE) allows us to extent the completeness of the RSs beyond the volume of the resonator, up to the boundary of the basis system.  The RSE rigorously calculates the unknow eigenmodes of a target system by using the complete set of know eigenmodes of a system as a basis for expansion \cite{MuljarovEPL2010}. It is capable of treating large perturbations of a resonator \cite{DoostPRA2014,LobanovPRA2019} and of the surrounding homogeneous medium \cite{AlmousaPRB2023}, with the accuracy and number of modes $N$ increasing with the maximum magnitude $\kmax$ of the complex wave numbers of basis modes included in the expansion \cite{DoostPRA2014,SztranyovszkyPRA2022}.

The purpose of this Letter is twofold. Firstly, we show that when the modes of a resonator are generated with the RSE, they form an asymptotically complete set over the basis system volume. Particularly, this means that in case of an embedded target system having a smaller volume, we can obtain an asymptotically complete set of its modes, and ML series of the GF 
%\cite{LeungPRA1994,LeungJOSAB1996,MuljarovEPL2010, WeissPRB2018, BothSST2021}
also converges to the correct values beyond the boundaries of the resonator of interest. Secondly, using the asymptotic completeness of the RSs inside the basis system, we demonstrate the applicability of the scattering theory \cite{LobanovPRA2018} to non-spherical resonators and, more generally, to resonators with boundaries different from the boundary of a basis spherical system. Previously, the scattering theory \cite{LobanovPRA2018}, based on the RSE and the link between the dyadic GF and the S matrix, was demonstrated for spherical systems only (and with material perturbations only), where the angular momentum quantum number $l$, magnetic quantum number $m$, transverse-electric (TE) and transverse-magnetic (TM) polarizations of light are conserved and can thus be separated. Here, we demonstrate the applicability of the theory to systems with cylindrical symmetry mixing the basis modes of different $l$ and polarizations.

\paragraph*{Completeness.}
\label{s:completeness}

First, we show that when the eigenmodes of a system are generated via the RSE, they form an asymptotically complete set over the basis system volume, which embeds the target system and thus includes the gap volume between of the basis and the target system boundaries. This principally is enabled by additional modes forming in the spectrum alongside the physical resonances in the system, and the additional modes depends on the basis size chosen. To illustrate this, we take a spherically symmetric dielectric system, which can be modeled as an effective one-dimensional (1D) problem, with eigenmodes being functions of the radial coordinate $r$ only. The unperturbed basis system has radius $R$, and we choose a the perturbed target system of radius $R_p=0.7R$, keeping the same permittivity. We calculate the modes in the complex wave number plane with the RSE, and then construct the GF as a ML series over the perturbed eigenmodes, both inside the perturbed resonator ($r<R_p$) and outside of it ($R_p<r<R$).

For simplicity we focus on the TE modes, for which the GF, having in general a dyadic form in 3D, can be represented in vector spherical harmonics (VSHs) by only one non-zero component and has no contribution from static modes \cite{MuljarovPRA2020}. We consider a non-magnetic, non-dispersive, homogeneous dielectric sphere in vacuum as background, for which the electric field $\mathcal{E}(r)$ satisfies the differential equation \cite{MuljarovPRA2020, SztranyovszkyPRA2022}
\begin{equation}
    \label{eq:radial_equation_TE}
    \left(  \dv[2]{}{r} - \frac{l(l+1)}{r^2} + k_n^2 \varepsilon(r)  \right) \mathcal{E}_n(r) = 0 \;,
\end{equation}
where $\varepsilon(r)$ is the permittivity, $k$ is the complex wave number, and $n$ labels different modes. The basis system has $\varepsilon(r) = \varepsilon_s - \Theta(r-R)(\varepsilon_s -1)$, and the target system $\varepsilon(r) = \varepsilon_s - \Theta(r-R_p)(\varepsilon_s -1)$, where $\Theta$ is the Heaviside step function, and we set the permittivity of the sphere to $\varepsilon_s=9$ as in \Onlinecite{LobanovPRA2018, LobanovPRA2019}.

\begin{figure}
    \centering
    \includegraphics[width=\linewidth]{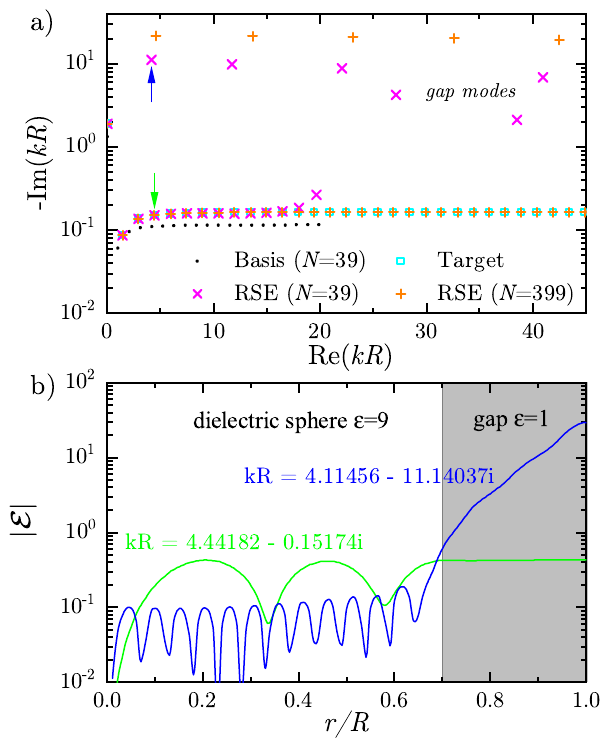}
    \caption{a) RSE modes (crosses)  of a dielectric sphere with $\varepsilon=9$, basis radius $R$, and target radius $0.7R$, in the complex wave number plane, along with the exact modes of the basis (dots) and target system (squares). b) Field amplitude of a mode close to a physical RS (green) and a gap mode (blue), with arrows in a) indicating these modes for $N=39$. }
    \label{fig:shell_modes}
\end{figure}

The modes of the basis (dots) and the target system (squares) are shown in \Fig{fig:shell_modes}a, alongside with modes calculated with the RSE (crosses) using two different basis sizes $N=39$ ($\kmax R \approx 20$) and $N=399$ ($\kmax R \approx 210$), for $l=1$, see \Onlinecite{MuljarovEPL2010, MuljarovPRA2020, SztranyovszkyPRA2022} or %Sec.\,\ref{S-s:rse} 
Sec. S.I of the Supplementary Information \cite{SI} for a detailed description of the RSE for this system. We can see that the target modes are correctly calculated via the RSE for wave numbers well below $\kmax$, as previously found. Increasing $\kmax$ has two effects: increase of the accuracy of modes that were already in the basis range and increase of the range where modes are approximately correctly found \cite{DoostPRA2012}. Looking at the results for $N=39$, we see that apart from modes corresponding to the exact target modes, there are additional modes present in the spectrum, with significantly larger imaginary part than the exact target modes, even an order of magnitude larger than the single \textit{leaky mode} \cite{SztranyovszkyPRA2022} appearing for $l=1$ on the imaginary axis (for modes with higher $l$ see %Sec.\,\ref{S-s:l10} 
Sec. S.II of \cite{SI}). As such, they have a small amplitude of $\mathcal{E}$ inside the target resonator while growing quickly outside it, as illustrated in \Fig{fig:shell_modes}b. These modes behave as if their fields were reflected from the surface of the basis resonator, which is evidenced by their spectral separation corresponding to the field quantization in the gap region.  We will therefore refer to these as \addmodes{} in the remainder of the paper.

We note that the classification of modes as gap modes can be done based on their spectral distance to the exact modes of the target system. However, the important  point here is that using the RSE, only modes with wave numbers well below $\kmax$ have small errors, leaving additional modes (typically up to a half of all modes) at larger wave numbers, not corresponding to exact modes. These modes are essential for the correct description of the system response. As we increase the basis size, the majority of these modes tend to infinite $k$ along the imaginary axis, rather than converge to the true physical modes (see \Fig{fig:shell_modes}a), thus they are clearly distinguishable, except for a few \textit{edge modes} at $k\approx\kmax$. The distance between the real parts of the wave numbers of gap modes  is about $\pi/(R-R_p)$, providing over the gap region a phase difference of $\pi$ between the neighboring gap modes. Consequently, their number for a given \kmax\ grows linearly with the gap size, as further detailed in %Sec.\,\ref{S-s:number_of_shell}
Sec. S.III of \cite{SI}. The effect of these modes on the asymptotic completeness of target modes is considered below, where we construct the ML series of the GF of the system.

Following the notation of \Onlinecite{MuljarovPRA2020}, the GF $\mathcal{G}$ corresponding to \Eq{eq:radial_equation_TE} satisfies the equation
\begin{equation}
    \label{eq:radial_gf_TE}
    \left(  \dv[2]{}{r} - \frac{l(l+1)}{r^2} + k^2 \varepsilon(r)  \right) \mathcal{G}(r,r';k) = k\delta(r-r') \;,
\end{equation}
and can be written as a ML series inside the resonator as
\begin{equation}
    \label{eq:radial_gf_ML_quick}
    \mathcal{G}(r,r';k) = \sum_n \frac{k\mathcal{E}_n(r) \mathcal{E}_n(r')}{k_n(k-k_n)} \,,
\end{equation}
or
\begin{equation}
    \label{eq:radial_gf_ML_slow}
    \mathcal{G}_{\mathrm sr}(r,r';k) = \sum_n \frac{\mathcal{E}_n(r) \mathcal{E}_n(r')}{k-k_n} \;,
\end{equation}
where the two equations are related by the sum rule
\begin{equation}
    \label{eq:radial_sum_rule}
    \mathcal{S} = \sum_n \frac{\mathcal{E}_n(r) \mathcal{E}_n(r')}{k_n} = 0 \;,
\end{equation}
as
%\begin{equation}
%    \label{eq:radial_gf_sum_rule}
%    \mathcal{G}_{\mathrm sr}(r,r';k) = \mathcal{G}(r,r';k)+ \frac{1}{k} \sum_n \frac{\mathcal{E}_n(r) \mathcal{E}_n(r')}{k_n} \;,
%\end{equation}
$\mathcal{G}_{\mathrm sr}(r,r';k) = \mathcal{G}(r,r';k)+ \mathcal{S}$.

\begin{figure}
    \centering
    \includegraphics[width=\linewidth]{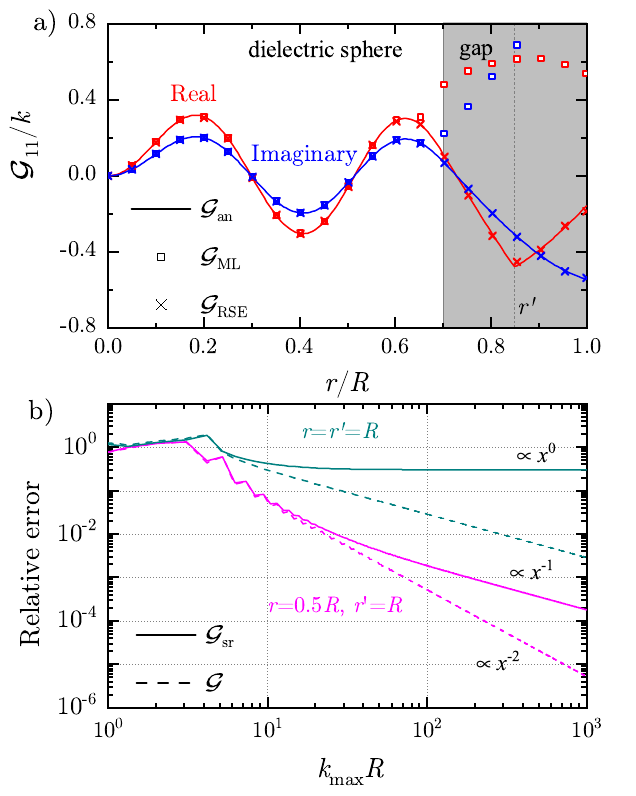}
    \caption{a) Comparison of the exact analytic form of the GF ($\mathcal{G}_\mathrm{an}$) and its ML series using exact modes ($\mathcal{G}_\mathrm{ML}$) with $N=39$ ($\kmax R\approx28.5$), and modes calculated via the RSE ($\mathcal{G}_\mathrm{RSE}$) with $N=39$ ($\kmax R \approx 20$), for a source located in the gap at $r'=0.85R$ (vertical dashed line), for $kR=5$. b) Relative error of the ML series of the GF, with one point and both points on the surface of the basis sphere, as labelled, for $kR=5$, calculated via \Eq{eq:radial_gf_ML_slow} (solid) and \Eq{eq:radial_gf_ML_quick} (dashed).}
    \label{fig:gf_outside}
\end{figure}

Now we evaluate the ML series of the GF using modes with wave numbers up to \kmax\ and compare it with the exact GF which can be found from two solutions of the homogeneous wave equation, satisfying the left and right boundary conditions separately \cite{RileyBook2006}; for spherically symmetric systems, a derivation can be found in \Onlinecite{MuljarovPRA2020}. For illustration, we place the source in the gap region at $r'=0.85R$ and show the results in \Fig{fig:gf_outside}a.

First let us consider the ML series over the exact target modes (squares). We can see that well inside the resonator ($r\lesssim0.6R$) the exact modes are sufficient to reproduce the GF despite one coordinate ($r'$) being located outside the resonator. In fact, when one coordinate is outside, but near the boundary, and one coordinate is inside, away from the boundary, a correct ML expansion using the exact eigenmodes is still possible, due to the analytic GF vanishing for $k\to \infty$ (which is always the case for both coordinates located inside the resonator \cite{LeungPRA1994, LeungJOSAB1996}), consistent with findings in \Onlinecite{KristensenAOP2020}, where the same was observed for a slab. If both coordinates are outside, the analytic GF is no longer vanishing for $k\to \infty$, and thus it cannot be written as a correct ML series using the exact eigenmodes only.

Now let us consider what happens when we use the modes generated via the RSE including the gap modes in the ML series (crosses). We can see that the GF is reproduced correctly across the whole unperturbed system volume, up to $R$, thus we can conclude that the gap modes dominantly contribute to the completeness in the gap, as also conformed by a detailed analysis in %Sec.\,\ref{S-s:excluding_modes} 
Sec. S.VII of \cite{SI}. This is also consistent with the fact that the perturbed modes $\mathcal{E}(r)$ are obtained via a rotation of the basis modes $\mathcal{E}_n(r)$, expressed by  the expansion $\mathcal{E}(r) = \sum_n c_n \mathcal{E}_n(r)$, where the expansion coefficients $c_n$ are given by a rotation matrix found from the RSE matrix equation (see %Sec.\,\ref{S-s:rse} 
Sec. S.I of \cite{SI}), and such a rotation preserves the completeness property within the volume of the basis system. Thus, generating eigenmodes via the RSE gives access to the GF also in the gap region (see %Sec.\,\ref{S-ss:ml_expansion_perturbed} 
Sec. S.IV.2 of \cite{SI} for a rigorous proof), and can provide more accurate results than any regularized or single-mode approximations of the GF \cite{DezfouliPRB2018, FrankePRA2023}.

We note that due to the $1/k_n$ factor in $\mathcal{G}(r,r';k)$, it has a quicker convergence than $\mathcal{G}_{\mathrm sr}(r,r';k)$, as illustrated in \Fig{fig:gf_outside}b. Such a difference comes from the sum rule \Eq{eq:radial_sum_rule} connecting both, which is exact only in the limit of an infinite, complete basis. Notably, $\mathcal{G}_{\mathrm sr}(r,r';k)$ does not even converge to the correct solution in the special case of both coordinates at the surface of the basis system, as illustrated in \Fig{fig:gf_outside}b, so the summation in $\mathcal{S}$ no longer tends to zero when $r=r'=R$, which is further discussed in %Sec.\,\ref{S-ss:sum_rule}  
Sec. S.IV.3 of \cite{SI}.
%We can also expect that there might be a particularly large error in $\mathcal{G}_{\mathrm sr}(r,r';k)$ around $k=0$, due to the $\mathcal{S}/k$ term, if the sum is truncated to a finite $n$.

\paragraph*{Scattering.}
\label{s:scattering}

\begin{figure}
    \centering
    \includegraphics[width=\linewidth]{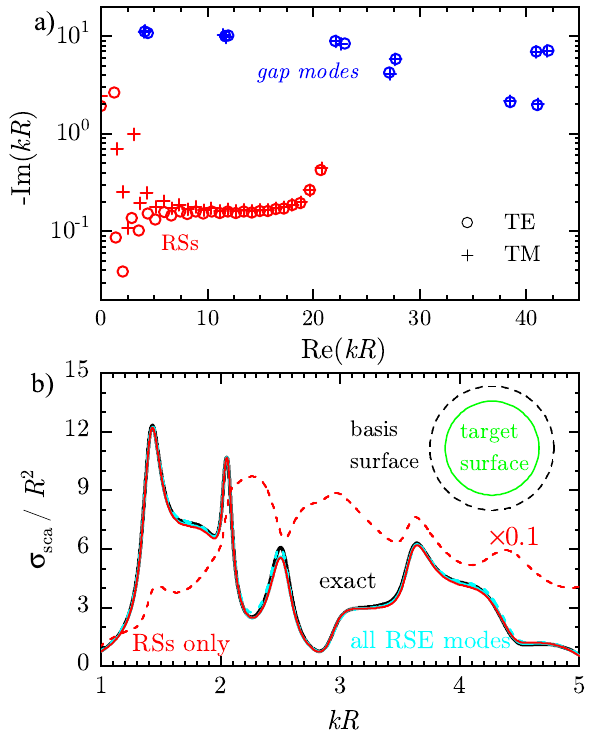}
    \caption{a) Complex $k$-plane with modes of a target dielectric sphere ($\varepsilon=9$) calculated with the RSE for $l=1,2$, both in TE and TM polarizations, with a basis $\kmax R \approx 20$ ($N=158$). b) Scattering cross-section of a perturbed sphere of radius $R_p$, with basis modes from a), calculated on the perturbed (solid lines) and unperturbed surface (dashed lines), with (teal) and without (red) gap modes.}
    \label{fig:scat_sph}
\end{figure}

We further illustrate the completeness by calculating the scattering cross-section of spherical resonators based on the GF evaluated on their surface, and away from the surface of the target resonator, but inside the basis system, including its surface. The importance of gap modes is exemplified in the latter case. We apply the method developed in \Onlinecite{LobanovPRA2018}, which is reviewed in %Sec.\,\ref{S-s:g-s} 
Sec. S.V of \cite{SI}, and is briefly summarised here: (i) the incoming excitation is decomposed into VSHs and TE and TM polarizations; (ii) a source term representing the excitation is found on a spherical surface surrounding the resonator; (iii) using the source term, the response of the resonator is found from the GF evaluated on this spherical surface; (iv) from this response, the S matrix and the scattering cross-sections are determined.
A fundamental result of the method is the link between the S matrix and the dyadic GF \cite{LobanovPRA2018}, which is given by
\begin{equation}
    \label{eq:s_matrix}
    S_{lmp}^{l'm'p'} = G_{lmp}^{l'm'p'}(R,R;k) \sigma_{l',p'}(k) - \delta_{pp'} \delta_{ll'} \delta_{mm'} \,,
\end{equation}
where $\delta$ is the Kronecker delta, $G_{lmp}^{l'm'p'}(R,R;k)$ are components of the radial part of the GF of Maxwell's wave equation for the electric field in the VSH representation, evaluated at the surface of a sphere of radius $R$, while $l$, $m$, and $p$ label, respectively, the VSH channels and polarization of the outgoing waves, with $l'$, $m'$, $p'$ denoting the incoming channels. We note that for a spherically symmetric system, $l$, $m$, and $p$ are not mixed and $G(R,R;k) = \mathcal{G}(R,R;k)/(kR^2)$, see %Sec.\,\ref{S-s:gf_link} 
Sec. S.VI of \cite{SI} for details. The effective source term $\sigma_{l',p'}(k)$ has a simple analytic form due to the use of spherical waves and is evaluated at the wave number $k$ of the incoming field. Using the S matrix, we can find the amplitude of the scattered field for each channel $(l,m,p)$, which then allows us to obtain the scattering cross-section $\sigma_{\rm sca}$, as detailed in %Sec.\,\ref{S-s:g-s} 
Sec. S.V of \cite{SI}. Importantly, the non-resonant background contribution to the S matrix and the cross-section are fully contained in \Eq{eq:s_matrix} via the Kronecker delta term and the static pole of the GF, and is not omitted as in \Onlinecite{BaiOE2013}, nor require further fitting as done in  \Onlinecite{WeissPRB2018}, nor approximations as used in \Onlinecite{RuanPRA2012}. Note that due to the convergence properties discussed above, we need to use the ML series \Eq{eq:radial_gf_ML_quick} for the S matrix in \Eq{eq:s_matrix}, to ensure its convergence to the correct value, whereas the RSE uses the ML series in the form of \Eq{eq:radial_gf_ML_slow} to obtain a linear eigenvalue problem.

We first consider the spherical target system treated before and calculate the cross-section on its surface, using $G_{lmp}^{l'm'p'}(R_p,R_p;k)$ in \Eq{eq:s_matrix}, and on the surface of the basis sphere, using $G_{lmp}^{l'm'p'}(R,R;k)$ in \Eq{eq:s_matrix}. We note that for a spherically symmetric system, the GF is diagonal, that is, the different channels $(l,m,p)$ do not mix leaving only $G_{lmp}^{lmp}$ non-zero. The cross-section is calculated with and without gap modes. The results are shown in \Fig{fig:scat_sph}(b), with the modes shown in \Fig{fig:scat_sph}(a).

When the cross-section is calculated on the surface of the perturbed sphere the results with and without gap modes are nearly identical. In this case there is no significant benefit or drawback from using these artificial modes in the calculation.
When the cross-section is calculated on the original, unperturbed surface, which now lies outside the perturbed resonator, the impact of the gap modes  is significant. If only the physical RSs are used in the calculation, the resulting cross-section is an order of magnitude larger than the exact solution and of largely different shape. This is expected, as using the RSs only, it is not possible to express the GF as a ML series with coordinates taken outside the resonator.
Including the gap modes in the calculation restores the convergence of the ML series, and the results show good agreement with the exact solution. This implies that the scattering formalism developed in \Onlinecite{LobanovPRA2018} should be suitable in case of non-spherical perturbation as well, where the spherical surface on which $G$ is evaluated can only be chosen outside and away from the resonator, as we will show below. The effect of excluding selected RSs or gap modes is further investigated in %Sec.\,\ref{S-s:excluding_modes} 
Sec. S.VII of \cite{SI}, with results showing that RSs that are far away from the spectral range of interest have small impact on the accuracy, while all gap modes, particularly in the spectral range of interest, have a significant impact, and they should not be neglected. 
We note that in this formalism, the static modes, which can be used to represent the static pole of the GF, are crucial for obtaining the correct S matrix and cross-section for non-dispersive dielectric systems as opposed to other methods \cite{SauvanOE2021, BesbesM2022}. This is due to the quantities being derived here directly from the GF, which itself has contributions from static modes, the latter adding non-resonant parts to the S matrix and cross-sections. Leaving out the static mode contribution leads to a considerable systematic error, as shown in %Sec.\,\ref{S-s:excluding_modes} 
Sec. S.VII of \cite{SI}. On the other hand, using the RSE allows us to calculate both the RSs and static modes together in the same matrix equation, and therefore does not require a separate static mode solver as in \Onlinecite{BesbesM2022}.

\paragraph*{Non-spherical systems.}

\begin{figure}
    \centering
    \includegraphics[width=\linewidth]{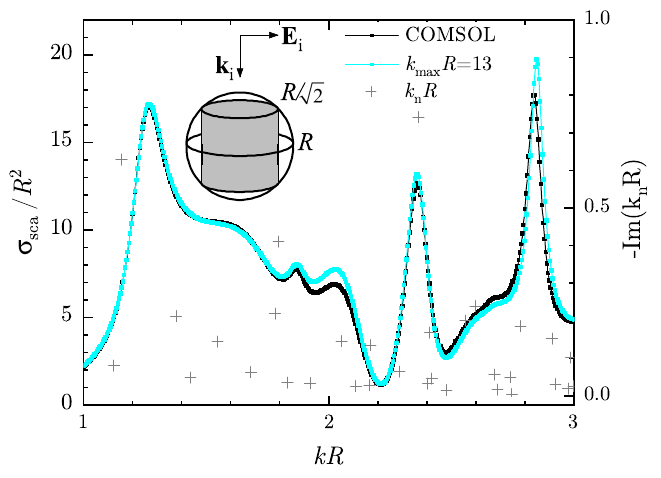}
    \caption{Scattering cross-section of a cylinder, with $\varepsilon=9$, height and diameter of $\sqrt{2}R$, and incoming excitation propagating along the cylinder axis (${\bf k}_{\rm i}$), calculated with $\kmax R = 13$ ($N=1004$). The eigenmodes in the complex plane are shown for comparison (crosses, right axis).}
    \label{fig:scat_cyl}
\end{figure}

We now demonstrate the applicability of the scattering theory when the spherical symmetry of the system is broken. We apply the scattering calculation method to a cylinder of equal diameter and height, $\sqrt{2}R$, with $\varepsilon=9$, as in \Onlinecite{MuljarovPRB2016Purcell, LobanovPRA2019}. The modes of the cylinder are calculated with the RSE (see %Sec.\,\ref{S-s:rse}
Sec. S.I of \cite{SI} for details) using the RSs of a sphere as basis. The results are shown in \Fig{fig:scat_cyl} and compared to a simulation with COMSOL Multiphysics\textsuperscript{\tiny\textregistered} \cite{COMSOL}. %The estimated accuracy of the COMSOL result is within $1\%$ of the exact value, based on the convergence of the amplitude of the scattering cross-section at the dipole peak around $k_dR = 1.25$ with respect to refining the size of mesh elements in the domain. 
%We can see that the dipole peak is reproduced within the error of the COMSOL calculation, already for $\kmax R = 6.5$, only five times $k_dR$. Higher order mode positions show a slight blueshift compared to COMSOL and usually an increase in amplitude. Both of these deviations are scaling approximately with $1/\kmax$. The convergence of the peak position corresponds to the convergence of the eigenfrequencies, as discussed in Sec.\,\ref{S-s:convergence} of \cite{SI}. The convergence of the scattering cross-section is due to the $1/\kmax$ convergence of the ML series for the relevant components of the dyadic GF (see \cite{MuljarovPRA2020} and Sec.\,\ref{S-ss:sum_rule} of \cite{SI}), and it is consistent with the observations for spherically symmetric systems \cite{LobanovPRA2018}. 
%As an alternative to increasing the number of modes for better overall accuracy, one could improve on the efficiency of the RSE for non-spherical systems by reformulating the static mode contribution, just as it was done for spherical systems in \Onlinecite{MuljarovPRA2020}, or could apply an extrapolation method \cite{DoostPRA2012} to increase the accuracy of the eigenfrequencies, thus the position of the peaks, or treat the increasing basis size in first order only, improving the accuracy of the amplitude of the peaks \cite{LobanovPRA2018}, as these have a lower computational complexity.
We can see good agreement between the two results, which further confirm completeness of the perturbed modes on the original surface of the basis sphere, outside of the cylidnrical resonator. A convergence study of the RSE results can be found in %Sec.\,\ref{S-s:convergence} 
Sec. S.VIII of \cite{SI}, showing that the accuracy improves as more modes are included in the basis.

\paragraph*{Conclusion.}
\label{s:conclusion}

We have shown that the eigenmodes of a target system generated by the RSE form an asymptotically complete set for ML expansion of the GF over the chosen basis system volume embedding the target system, which includes a gap volume surrounding it. Completeness in the gap volume is achieved by including \addmodes{} in the spectrum alongside the RSs and static modes of the target system. This allows one to calculate the response to a source in the gap volume, outside the target resonator. As a result, we find that the scattering theory which links the GF to the S matrix \cite{LobanovPRA2018} is accurate for a spherical target system, with the spherical scattering interface taken at the surface of either the basis or the target system. Importantly, we extend the application of this scattering theory to non-spherical systems, namely to a cylinder, supporting its validity for arbitrary target systems.

The provision of a complete set of eigenmodes for ML expansion of the GF of an open system in a volume including an adjustable gap region surrounding the system enables important applications in the simulation of open resonators for sources, sensors, and detectors of radiation.

\begin{acknowledgments}
 Z.S. acknowledges the Engineering and Physical Sciences Research Council for his PhD studentship award (grant EP/R513003/1).
\end{acknowledgments}

\bibliographystyle{unsrt}
\bibliography{refs}

\end{document}